\documentclass[12pt,leqno]{amsart}

\usepackage{amsmath,amssymb,amscd,amsthm}
\usepackage{latexsym}
\usepackage{epsfig}

\newtheorem{theorem}{Theorem}
\newtheorem{lemma}{Lemma}
\newtheorem{proposition}{Proposition}

\newtheorem{corollary}{Corollary}

\newtheorem{claim}{Claim}

\newcommand{\q}{\quad}
\newcommand{\qq}{\quad\quad}

\newcommand{\norm}[2]{{\left\| #1 \right\|}_{#2}}
\newcommand{\f}[2]{\frac{#1}{#2}}
\newcommand{\dpr}[2]{\langle #1,#2 \rangle}

\newcommand{\al}{\alpha}

\newcommand{\De}{\Delta}
\newcommand{\ve}{\varepsilon}

\newcommand{\La}{\Lambda}
\newcommand{\si}{\sigma}

\newcommand{\vp}{\varphi}

\newcommand{\rone}{\mathbf R^1}

\newcommand{\cz}{\mathcal Z}

\newcommand{\intl}{\int\limits}

\newcommand{\suml}{\sum\limits}
\newcommand{\supl}{\sup\limits}

\newcommand{\p}{\partial}

\newcommand{\beq}{\begin{equation}}
\newcommand{\eeq}{\end{equation}}
\newcommand{\beqna}{\begin{eqnarray*}}
\newcommand{\eeqna}{\end{eqnarray*}}
\newcommand{\beqn}{\begin{equation*}}
\newcommand{\eeqn}{\end{equation*}}
\newcommand{\bp}{\begin{proof}}
\newcommand{\ep}{\end{proof}}
\newcommand{\bprop}{\begin{proposition}}
\newcommand{\eprop}{\end{proposition}}
\newcommand{\bt}{\begin{theorem}}
\newcommand{\et}{\end{theorem}}
\newcommand{\bex}{\begin{Example}}
\newcommand{\eex}{\end{Example}}
\newcommand{\bc}{\begin{corollary}}
\newcommand{\ec}{\end{corollary}}
\newcommand{\bcl}{\begin{claim}}
\newcommand{\ecl}{\end{claim}}
\newcommand{\bl}{\begin{lemma}}
\newcommand{\el}{\end{lemma}}

\begin{document}

\title
[Discrete Schr\"odinger/Klein-Gordon equation]
{Asymptotic behavior of small solutions for the discrete 
nonlinear Schr\"odinger and Klein-Gordon
equations }

\author{A. Stefanov}
\author{P.G. Kevrekidis}

\address{Atanas Stefanov\\
Department of Mathematics \\
University of Kansas\\
1460 Jayhawk Blvd\\ Lawrence, KS 66045--7523}

\address{Panayotis Kevrekidis\\
Lederle Graduate Research Tower\\ 
Department of Mathematics and  Statistics\\
University of Massachusetts\\
Amherst, MA 01003}

\thanks{The first author supported in part by  NSF-DMS 0300511 and the 
University of Kansas General Research Fund
\# 2301716. The second author is supported in part by NSF-DMS 0204585,
an NSF CAREER award and the Eppley Foundation for Research.}
\date{\today}

%\subjclass{ }

\begin{abstract}
We show  decay estimates for the propagator of the discrete Schr\"odinger
and Klein-Gordon equations in the form $\norm{U(t)f}{l^\infty}\leq C (1+|t|)^{-d/3}\norm{f}{l^1}$.
This implies a corresponding (restricted) set of Strichartz estimates.
Applications of the latter include  the existence of excitation thresholds  
for certain regimes of the parameters and the decay of small initial 
data for relevant $l^p$ norms. The analytical decay estimates  
are corroborated with numerical results.  
%(MORE STUFF HERE)
\end{abstract}

%\submitto{\NL}
\maketitle
\date{today}

\section{\bf Introduction}
A sequence of (time evolving) 
harmonic oscillators which interact only with their 
immediate
neighbors  is described by the discrete 
Schr\"odinger equation
$$
\left|\begin{array}{l}
iu_n'(t)+ h^{-2}(u_{n+h}(t)+u_{n-h}(t)-2u_n(t))+F_n(t)=0 
\q k\in h{\mathcal Z}\\
\{u_k(0)\}\in l^2
\end{array}
\right.
$$
where $h$ is the distance between the oscillators and $h{\mathcal Z}$ is the
lattice of points $\{h n: n - integer\}$ . 
More generally, 
$$
\Delta_{discrete}=h^{-2}\sum_{j=1}^d (u_{n+h e_j}+u_{n-h e_j}-2u_n)
$$
and the equation describing the system  is given by
$$
i u_n'(t)+ \Delta_{discrete} u+ F_n(t)=0, \q\q n\in h{\mathcal Z}^d,
$$
As $h\to 0$, one obtains the continuous model.\\
%(WE NEED MORE STUFF HERE)

For $F_n(t)=\pm |u|^{2 \sigma} u$, the above equation becomes
the discrete nonlinear Schr{\"o}dinger equation.
The latter  is one of the prototypical 
differential-difference models that is both
physically relevant and mathematically tractable.
Perhaps, the most direct implementation of this equation can be
identified in one-dimensional arrays of coupled optical waveguides
\cite{EMSAPA02,PMAAESPL02}. These may be multi-core structures
created in a slab of a semiconductor material (such as AlGaAs), or
virtual ones, induced by a set of laser beams illuminating a
photorefractive crystal. In this experimental implementation, there
are about forty lattice sites (guiding cores), and the localized,
solitary wave structures that this model is well known to support 
\cite{KRB01,EJ03} may propagate over tens of diffraction
lengths.

Photonic lattices \cite{ESCFS02,SKEA03} have
recently provided another application of the discrete 
nonlinear Schr{\"o}dinger class of models. In
this case, the refractive index of a nonlinear medium changes
periodically due to a grid of strong beams, while a weaker probe
beam is used to monitor the localized waves. This has created
a large volume of recent activity in the direction of understanding
discrete solitons in such photonic lattices.

Finally, besides its applications in nonlinear optics, the discrete
nonlinear Schr{\"o}dinger equation is a relevant model for 
Bose-Einstein condensates trapped in strong optical lattices
(formed by the interference patterns of laser beams) 
\cite{CBFMMTSI01,CFFFMI03}. In this context, the model can be
derived systematically by using the Wannier function expansions of
\cite{AKKS02}.

These applications render the discrete nonlinear Schr{\"o}dinger 
equation a particularly relevant dynamical lattice for a variety
of physical applications. 

Another example of a differential-difference
equation that arises in many physical
contexts consists of the nonlinear Klein-Gordon models.
%Discrete nonlinear Klein-Gordon models are also relevant to a host
%of physical applications. 
Perhaps the simplest possible implementation of such a lattice
arises for an array of coupled torsion pendula under the effect
of gravity \cite{pel}. However, such models are also relevant 
in condensed matter physics 
(e.g., describing the fluxon dynamics in arrays of 
superconducting Josephson-junctions)
\cite{cirillo}, as well as biophysics (e.g., describing the local
denaturation of the DNA double strand \cite{pb}). It is also
interesting to note that nonlinear Klein-Gordon models
are intimately related to their Schr{\"o}dinger siblings since
the latter are the natural envelope wave reduction
of the former \cite{kp}.

%\subsection{The Klein-Gordon model}
The Klein-Gordon model is of the form
\begin{equation}
\label{eq:200}
\left|\begin{array}{l}
 \p_t^2 u_n(t)- \Delta_{d} u+u_n+F_n(t)=0 \\
u_n(0)=f_n\in l^2(\cz^d), \\
\p_t u_n(0)=g_n \in l^2(\cz^d)
\end{array}
\right.
\end{equation}
where we take $h=1$ and the nonlinearity will in general be assumed
to be of the form
$F_n=\pm |u|^{2 \sigma} u$.
%the form $F_n=u_n^3$. \\

In this work, we examine such Schr{\"o}dinger and Klein-Gordon
models as follows: in section 2, we give a priori energy and decay
estimates for the free propagation in these lattices, which we
subsequently prove by means of Strichartz estimates in section 3.
In sections 4 and 5, we use these estimates to examine solutions
in the presence of nonlinearity for the Schr{\"o}dinger and Klein-Gordon
cases respectively. As an application, we show the existence of 
energy thresholds for the appearance of localized solutions 
(for appropriate regimes of the parameters) and
illustrate the decay of various lattice norms, complementing the
analysis with numerical simulations. A number of technical details
are worked out in the appendix.
\newpage
\section{Free Evolution: Decay and Energy Estimates}

\subsection{Schr\"odinger equation}
\begin{theorem}
\label{theo:1}
For the free discrete Schr\"odinger equation
$$
\left|\begin{array}{l}
i u_n'(t)+ \Delta_{d} u=0 \\
u_n(0)\in l^2(\cz^d)
\end{array}
\right.
$$
one has 
\begin{eqnarray}
& &
\label{eq:1}
\norm{\{u_n(t)\}}{l^2}=\norm{\{u_n(0)\}}{l^2}\qq \textup{energy identity} \\
& &
\label{eq:2}
\norm{\{u_n(t)\}}{l^\infty}\leq
C\frac{h^{2d/3}}{|t|^{d/3}}\norm{\{u_n(0)\}}{l^1}\qq \textup{decay estimate}
\end{eqnarray} 
Moreover,  for the inhomogeneous equation 
$$
\left|\begin{array}{l}
i u_n'(t)+ \Delta_{discrete} u_n +F_n(t)=0 \\
u_n(0)\in l^2
\end{array}
\right.
$$
one  has the Strichartz estimates with $(q,r)\geq 2,  
1/q+d/(3r)\leq d/6$ and \\ $(q,r,d)\neq (2,\infty, 3)$.  That is 
\begin{eqnarray*}
& &h^{2/q}\left(\intl_{0}^\infty \left(\suml_{n\in h\cz^d} |u_n(t)|^r\right)^{q/r} 
dt\right)^{1/q}\leq 
C\left(\suml_{n\in h\cz^d} |u_n(0)|^2\right)^{1/2}+\\
&+&C h^{2+2/\tilde{q}'}\left(\intl_{0}^\infty \left(\suml_{n\in h\cz^d} 
|F_n(t)|^{\tilde{r}'}\right)^{\tilde{q}'/\tilde{r}'} dt\right)^{1/\tilde{q}'}
\end{eqnarray*}
\end{theorem}
\noindent {\bf Remark} Note that the decay rate (and consequently the Strichartz estimates) is smaller than
the usual $t^{-d/2}$ that one has for the continuous analogue.

%%%%%%%%%%%%%%%%%%%%%%%%%%%%%%%%%%%%%%%

\begin{figure}[tbp]
\epsfxsize=8cm %\centerline{}
\centerline{\epsffile{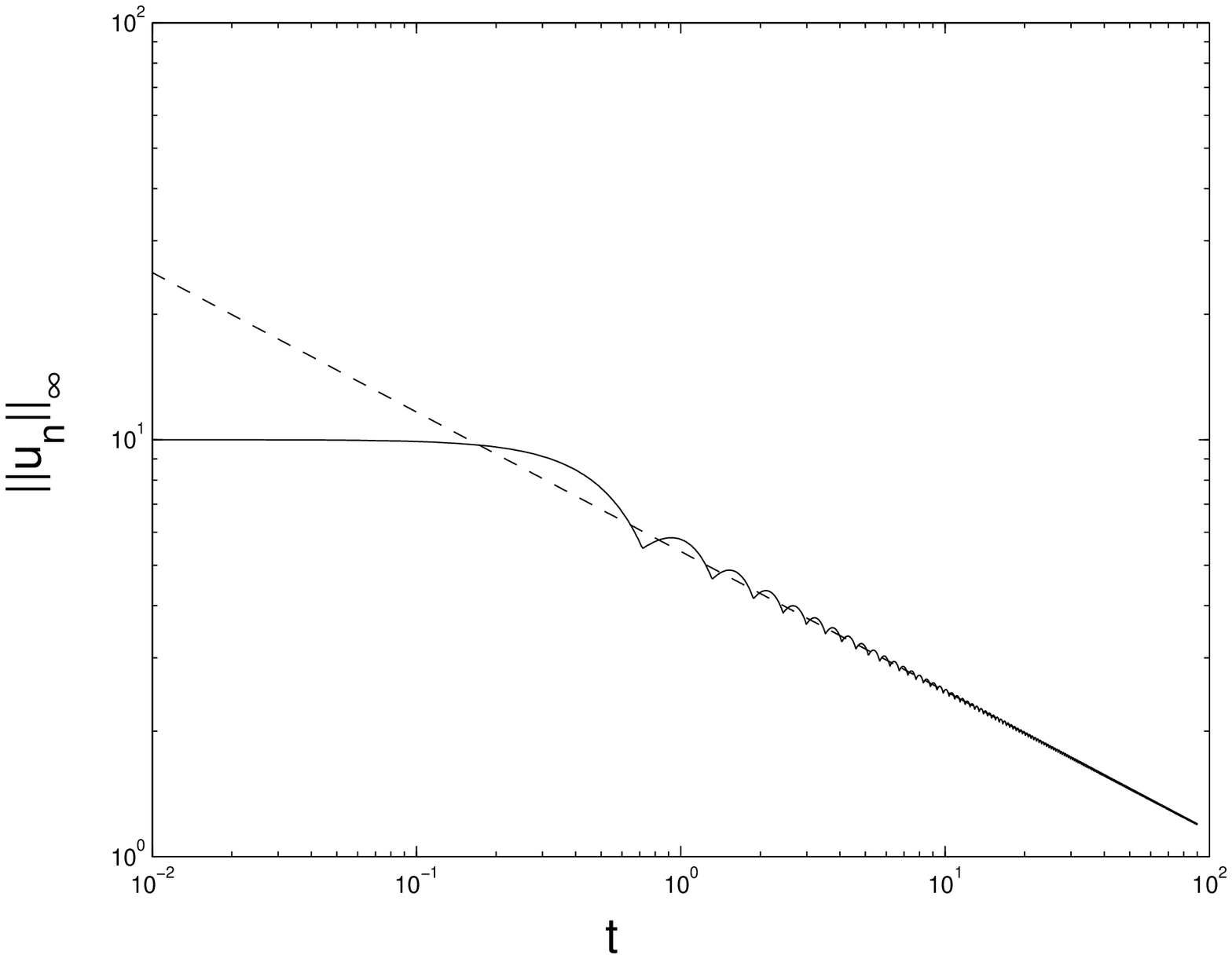}}
\caption{Log-log plot of the temporal evolution of the $L^{\infty}$ norm in a 
linear one dimensional 
Schr{\"o}dinger lattice. The intial condition contains one site excited
with $u_0=10$. The dashed line shows the best fit (for times 
$50 \leq t \leq 90$) $\sim t^{-0.335}$.}
\label{dfig1}
\end{figure}

%%%%%%%%%%%%%%%%%%%%%%%%%%%%%%%%%%%%%%%

In the Klein-Gordon case, similarly to 
the  Schr\"odinger models, we discuss various relevant estimates for the free
evolution. We have the following analogue of Theorem \ref{theo:1}:
\begin{theorem}
\label{theo:10}
For the solutions of the  one-dimensional\footnote{At this stage, we have encountered some technical difficulties, when trying to extend the result
to dimensions higher than one. In fact, the proof of the energy estimate \eqref{eq:201} goes through in all
dimensions, whereas the decay estimate \eqref{eq:202} boils down to the exact rate of decay of an explicit
d dimensional integral, see \eqref{eq:1200}. While numerically, one can see that the relevant
integral has the right rate of decay, its rigorous justification remains open.} homogeneous discrete 
  Klein-Gordon equation \eqref{eq:200}, one has the a priori decay and energy estimates 
\begin{eqnarray}
\label{eq:201}
& &\norm{\{u_n(t)\}}{l^2}\leq C( \norm{f_n}{l^2}+\norm{g_n}{l^2}),\q\textup{energy estimate}\\
\label{eq:202}
& &\norm{\{u_n(t)\}}{l^\infty}\lesssim t^{-1/3}  (\norm{f_n}{l^1}+\norm{g_n}{l^1})
\q\textup{decay estimate}\
\end{eqnarray}
For the solutions of the inhomogeneous equation, one has the Strichartz estimates with 
$(q,r)\geq 2,  
1/q+1/(3r)\leq 1/6$.  That is 
\begin{eqnarray*}
&  &\left(\intl_{0}^\infty \left(\suml_{n\in \cz} |u_n(t)|^r\right)^{q/r} 
dt\right)^{1/q}\leq 
C\left(\suml_{n\in \cz} |f_n|^2+|g_n|^2 \right)^{1/2}+\\
&+&C \left(\intl_{0}^\infty \left(\suml_{n\in \cz} 
|F_n(t)|^{\tilde{r}'}\right)^{\tilde{q}'/\tilde{r}'} dt\right)^{1/\tilde{q}'}
\end{eqnarray*}
\end{theorem}

%%%%%%%%%%%%%%%%%%%%%%%%%%%%%%%%%%%%%%%

\begin{figure}[tbp]
\epsfxsize=8cm %\centerline{}
\centerline{\epsffile{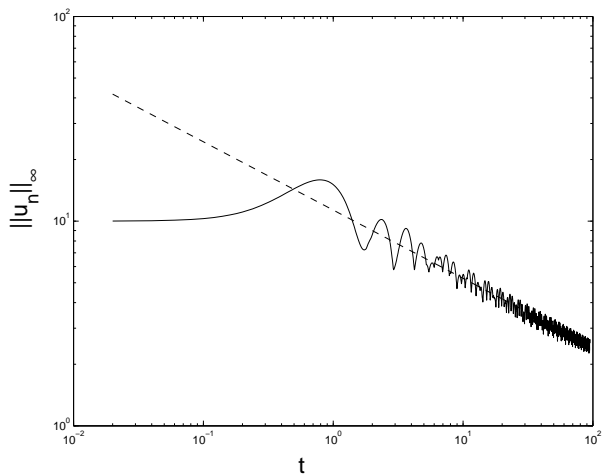}}
\caption{Log-log plot of the temporal evolution of the $L^{\infty}$ norm in a 
linear one dimensional Klein-Gordon 
lattice. The intial condition contains one site excited
with $u_0=10$. The dashed line shows the best fit (for times 
$20 \leq t \leq 90$) $\sim t^{-0.3335}$.}
\label{dfig1b}
\end{figure}

%%%%%%%%%%%%%%%%%%%%%%%%%%%%%%%%%%%%%%%

\section{Strichartz estimates for the Schr\"odinger and Klein-Gordon equations}
\noindent In this section, we present the proof of Theorem \ref{theo:1}. 
\begin{proof}(Theorem \ref{theo:1})
By rescaling in time, it suffices to consider the case $h=1$. Next, associate 
$\{u_n\}\longleftrightarrow f(k) =\suml_{n\in \cz^d} u_n e^{ 2\pi i n\cdot k}$. Then, 
$$
\Delta_{discrete}=\suml_{j=1}^d (S_j+S_j^*-2Id)
$$
where $S_j$ is the shift operator in the direction of $e_j$. On the spaces of trigonometric polynomials, 
$$
\Delta_{d} f= \suml_{j=1}^d (e^{2\pi ik_j}+e^{-2\pi ik_j }-2) f(x)=\left[-4\suml_{j=1}^d
\sin^2(\pi k_j) \right]f(k).
$$
Since $\norm{\{u_n\}}{l^2}=\norm{f}{L^2([0,1]^d)}$,  it is easy to see that 
$U(t)u_0=e^{- 4 i t \sum_{j=1}^d 
\sin^2(\pi k_j)}
f(k)$ is an isometry in $l^2$. \\
For the decay estimate \eqref{eq:2}, write 
$U(t)u_0 = e^{-4 i t \sum_{j=1}^d 
\sin^2(\pi k_j)} \left(\suml u_n(0) e^{ 2\pi i k\cdot n}\right)$, whence
\begin{equation}
\label{eq:203}
u_n(t)= \suml_{m\in \cz^d} u_m(0) \intl_{[0,1]^d} e^{-4 i t \sum_{j=1}^d
\sin^2(\pi k_j)} e^{2\pi i (m-n)\cdot k}dk.
\end{equation}
Thus, \eqref{eq:2}, reduces to showing 
$$
\supl_{m,n\in \cz^d}\left|\intl_{[0,1]^d} e^{-4 i t \sum_{j=1}^d
\sin^2(\pi k_j)} e^{2\pi i (m-n)\cdot k}dk\right|\leq C \min(1, |t|^{-d/3}).
$$
Since the integrals split into 1 D integrals, it suffices to show (after elementary change of variables)
$$
\supl_{s\in \rone}\left|\intl_0^{1} e^{- i t( \sin^2(x)-s x)}dx\right|\leq C \min(1, |t|^{-1/3}).
$$
The estimate by $1$ is trivial by taking absolute values, while the estimate by $|t|^{-1/3}$ follows from
the Van der Corput lemma, see \cite{S}, p. 332. Indeed, we have to verify that the phase function 
$\vp(x)=\sin^2(x)-s x$ satisfies \\
$\max(|\vp'(x)|, |\vp''(x)|, |\vp'''(x)|)\geq 1$, which is verified since
$(\vp'')^2(x)+(\vp''')^2(x)=4(\cos^2(2x)+\sin^2(2x))=4$. \\
Note that if $s=1$, we may have $\max(|\vp'(x)|,
|\vp''(x))|<<1$ for $x\sim \pi/4$, indicating that this is the sharp rate of decay, at least as far as 
the Van der Corput lemma is concerned.\\
Corroborating this estimate in a direct numerical simulation,
we have considered a linear one-dimensional lattice with 
$h=1$. The result is shown in Fig. \ref{dfig1} for the temporal
evolution of the $L^{\infty}$ norm. The best fit of the 
numerical result in the log-log plot of the figure very closely
follows the theoretical prediction since the corresponding decay
exponent is found to be $\approx 0.335$.

The Strichartz estimates follow from the abstract Strichartz estimate by 
 Keel and Tao, \cite{KeelTao}, which states that
energy and decay estimates imply Strichartz estimates. The last statement is also implicit in the earlier
work of Ginibre and Velo, \cite{Ginibre}.
\end{proof}

The proof of the Strichartz estimate for the discrete Klein-Gordon equation (DKG) follows closely the
proof for the Schr\"odinger case. There are however some distinctive differences, which we try to
highlight in the argument.
\begin{proof}(Theorem \ref{theo:10})
Let us consider  the homogeneous equation
first.\\ 
The energy estimate is derived in all dimensions. Let $\{u_n\}$ be a solution and 
multiply both sides of the (DKG) by $\p_t u_n(t)$ and sum in $n$. We have 
$$
\p_t \sum_n (\p_t u_n)^2 -\sum_n (\De_d u_n) \p_t u_n +\p_t \sum_n u_n^2 =0\\
$$
By the summation by parts formula, (e.g. (2.4) in \cite{weinstein}, it is not hard to see that
%(CHECK THAT FORMULA PLEASE)  that 
% PGK: it's not hard to check the formula since it yields the energy identity
% hence it has to be correct.
$$
-\sum_n (\De_d u_n) \p_t u_n = \p_t \suml_{r=1}^d \sum_n |u_n-u_{n+e_r}|^2/2.
$$ 
Thus, 
$$
\p_t (\sum_n (\p_t u_n)^2+\suml_{r=1}^d \sum_n |u_n-u_{n+e_r}|^2+\sum_n u_n^2)=0, 
$$
implying that 
\begin{eqnarray*}
\sum_n u_n^2(t) 
 &\leq& (\sum_n (\p_t u_n)^2+\suml_{r=1}^d \sum_n |u_n-u_{n+e_r}|^2+\sum_n u_n^2)=\\
 &=& (\sum_n g_n^2+\suml_{r=1}^d \sum_n |f_n-f_{n+e_r}|^2+\sum_n f_n^2)
 \leq C (\sum_n g_n^2+\sum_n f_n^2),
\end{eqnarray*}
which is \eqref{eq:201}. \\ 
We have already introduced and studied the action of $\De_{disc}$ (where for simplicity 
$\De_{disc}=\De_1$ is the
discrete Laplacian in 1 D),  see \eqref{eq:203}. Following the same idea,
consider $\{u_n\} \longleftrightarrow f=\sum_n u_n e^{2\pi n\cdot k}$. Then, 
from Duhamel's formula
for the wave equation (and also by straightforward verification),
the solution to the inhomogeneous equation is given by 
\begin{eqnarray*}
u_n(t)=\cos(t \sqrt{1-\De_{disc}}) f_n+\f{\sin (t \sqrt{1-\De_{disc}})}{\sqrt{1-\De_{disc}}} g_n + 
\intl_0^t \f{\sin ((t-s) \sqrt{1-\De_{disc}})}{\sqrt{1-\De_{disc}}} F_n(s) ds.
\end{eqnarray*}
where $\sin(x), \cos(x)$ are expressed via the Euler's formula in terms of 
$e^{i t x}$ and a (continuous) function of $\sqrt{1-\De_{disc}}$ is the operator 
  given by 
\begin{eqnarray*}
(h(\sqrt{1-\De_{disc}})u)_n(t)=\suml_{m\in\cz} 
u_m(0) \intl_0^1 h(1+4\sin^2(\pi k)) e^{2\pi i(m-n) k}dk.
\end{eqnarray*}
We have the following technical lemma, whose  proof  is postponed for the appendix.
\begin{lemma}
\label{le:1}
The operator $h(\sqrt{1-\De_{disc}}):l^p\to l^p$ for all $1\leq p\leq \infty$ for all sufficiently
smooth functions $h$.
\end{lemma}
By the abtract result of Keel and Tao, the energy estimate \eqref{eq:201} and Lemma \ref{le:1}, 
matters reduce
to establishing the decay estimate 
\begin{equation}
\label{eq:205}
\norm{e^{i t \sqrt{1-\De_{disc}}}u}{l^\infty}\leq C t^{-1/3} \norm{u_n}{l^1}.
\end{equation}
Indeed, for Duhamel's term, we have by the above mentioned 
result of Keel and Tao
\begin{eqnarray*}
& &\norm{\intl_0^t \f{\sin ((t-s) \sqrt{1-\De_{disc}})}{\sqrt{1-\De_{disc}}} F_n(s) ds}{L^ql^r}\lesssim 
\norm{(1-\De_{disc})^{-1/2} F}{L^{q'}l^{r'}}\lesssim \norm{F}{L^{q'}l^{r'}},
\end{eqnarray*}
where the last inequality follows from Lemma \ref{le:1}.

\noindent
As in the proof of \eqref{eq:203}, one needs to show 
\begin{equation}
\label{eq:206}
\supl_{m\in \cz}\intl_0^1 e^{ i t\sqrt{1+4\sin^2(\pi k )}} e^{-2\pi i m k}dk\leq C t^{-1/3},
\end{equation}
which by an elementary change of variables would follow from 
\begin{equation}
\label{eq:806}
\supl_{s\in\rone} \intl_0^1 e^{ i t(\sqrt{1+\sin^2(x )}- s x)} dx \leq C t^{-1/3},
\end{equation}
Thus,  the phase function is $\psi(x)=
\sqrt{1+\sin^2(x)}-s x$ and let us denote the function $\vp(x)=\sqrt{1+\sin^2(x)}$. Denote
$\al=\arccos(2-\sqrt{2})$. 

\noindent 
We have that the solutions to $\psi''(x)=0$ in $[0,2\pi]$ are 
 $x=\al, \pi-\al, \pi+\al, 2\pi-\al$. As we know by the Van der Corput lemma, the worst rate of decay 
 occurs, when  some consecutive derivatives (starting from $\psi'$) vanish 
 at a point. In our case,
 we have that only for $s=\vp'(\al), \vp'(\pi-\al), \vp'(\pi+\al), \vp'(2\pi-\al)$, one has
 $\psi'(x)=\psi''(x)=0$ ( at the points $x=\al, \pi-\al, \pi+\al, 2\pi-\al$). It is easy to check 
%(say by
% using Mathematica), 
that for these 
values of $s$, one has $\max(\psi'(x), \psi''(x), \psi'''(x))\geq
 1/2$, which guarantees by the Van der Corput lemma a decay of at least $t^{-1/3}$.
% (PLEASE SHOW THE SHARPNESS OF THIS RATE NUMERICALLY)
% PGK: this is not very straightforward numerically. I think what is best
% to do here is to try to show the same estimate as for the NLS for the 
% KG (in the applications section below).

\end{proof}
\noindent {\bf Remark} The proof of the higher dimensional analogue of such result  would involve showing that 
\begin{equation}
\label{eq:1200}
\supl_{s\in {\mathbf R}^d} \intl_{[0,1]^d} e^{ i t(\sqrt{1+\sum_{j=1}^d \sin^2(x_j)}- \dpr{s}{x})} dx \leq C t^{-d/3},
\end{equation}
Clearly, this integral does not reduce to the one dimensional case as in the Schr\"odinger case, which makes it
harder object to study. Numerical simulations seem to confirm the validity of \eqref{eq:1200} since in a two dimensional numerical experiment for a DKG
lattice, the best fit to the decay was found to be $\sim t^{-0.675}$. \\
%{\bf Do you want to try this Panos, I have tried it with Mathematica, it seems to work, 
%but you may want to try too}

\section{Applications to the discrete Schr\"odinger equation}
Consider 
\begin{equation}
\label{eq:150}
\left|\begin{array}{l}
i \p_t u_n+ \Delta_{d} u\pm |u_n|^{2\si} u_n=0 \\
u_k(0)\in l^2
\end{array}
\right.
\end{equation}
\subsection{Global solutions for small data in the regime $\si\geq 3/d$}
For \eqref{eq:150}, we establish the global existence of small solutions, provided 
$\norm{\{u_n(0)\}}{l^2}<<1$ and $\si\geq 3/d$. 
\begin{theorem}
\label{theo:2}
Let $\si\geq 3/d$. There exists an $\ve=\ve(d)>0$ and a constant $C=C(d)$, so that whenever  
$\norm{\{u_n(0)\}}{l^2}\leq \ve$, a unique global  solution to \eqref{eq:150} exists, which satisfies
$$
\norm{\{u_n(t)\}}{L^ql^r}\leq C\ve,
$$
for all Strichartz admissible pairs $(q,r)$. In particular, 
$\norm{u_n(t)}{l^r}\to 0$ as $t\to\infty$ for every $r>2$.
\end{theorem}
\begin{proof}
We perform an iteration procedure for the equation 
$$
u_n(t)=\La u_n=e^{i t \De_d} u_n(0)\pm i \intl_0^t e^{i (t-s) \De_d}|u_n|^{2\si} u_n(s) ds
$$ 
in the (metric) space $X=\{u: \supl_{(q,r)-admissible}\norm{u}{L^ql^r}<2 C \norm{u_n(0)}{l^2}\}$, where $C$ is the constant in the
Strichartz inequality, that is,  we are seeking  a fixed point of the map $\La$. 
Clearly, by the Strichartz estimates with
$(q,r)$ and $\tilde{q}'=1, \tilde{r}'=2$, we have 
\begin{eqnarray*}
& & \supl_{(q,r)-admissible}\norm{\La u_n(t)}{L^q l^r}
\leq C \norm{u_n(0)}{l^2}+ C_1\norm{|u_n|^{2\si+1}}{L^1l^2}\leq \\
&\leq& C
\norm{u_n(0)}{l^2}+C_1 \norm{u_n}{L^{2\si+1}l^{2(2\si+1)}}^{2\si+1} \leq  
C \norm{u_n(0)}{l^2}+C_1\norm{u_n}{X}^{2\si+1}\leq \\
&\leq & C \norm{u_n(0)}{l^2}+ C_1 (2C\ve)^{2\si+1}.
\end{eqnarray*}
In the above sequence of inequalities, we have used that since $\si\geq 3/d$, \\
$(2\si+1, 2(2\si+1))$
is a Strichartz admissible pair and therefore controllable by the norm in $X$.
Thus, for an appropriate absolute 
$\ve>0$, one has 
$$
\supl_{(q,r)-admissible}\norm{\La u_n(t)}{L^q l^r}\leq 2C\norm{u_n(0)}{l^2},
$$
that is $\La:X\to X$. \\
Similarly one verifies that 
\begin{eqnarray*}
\norm{\La (u_n(t)-v_n(t))}{L^q l^r}&\lesssim &
\norm{u_n-v_n}{L^{2\si+1}l^{2(2\si+1)}}(\norm{u_n}{L^{2\si+1}l^{2(2\si+1)}}^{2\si}+ 
\norm{v_n}{L^{2\si+1}l^{2(2\si+1)}}^{2\si})\lesssim \\
&\lesssim& \norm{u_n-v_n}{L^{2\si+1}l^{2(2\si+1)}}
\norm{u_n(0)}{l^2}^{2\si}.
\end{eqnarray*}
which by the smallness of $\norm{u_n(0)}{l^2}$ implies that the 
map $\La:X\to X$ is a contraction. This shows the existence of a global solution of 
$u=\La u$ and
by construction $\norm{u}{L^ql^r}<2 C \norm{u_0}{l^2}$.
\end{proof}

\subsection{Decay of small solutions and the Weinstein conjecture}
M. Weinstein has proved that for $\si\geq 2/d$, one has 
an energy excitation treshold,
\cite{weinstein}, i.e., 
that there exists $\ve=\ve(d)$, so that every standing wave solution 
$\{e^{i\La t} \phi_n\}$ must satisfy $\norm{\phi}{l^2}\geq \ve$. 
In the same paper, he has also conjectured that for sufficiently
small solutions, one has
$\lim_{t\to\infty} \norm{u(t)}{l^p}=0$ for all $p\leq \infty$. 

In the next theorem, we give conditions under which small solutions {\it will actually decay like the
free solution in the corresponding $l^p$ norms}. This of course is a statement implying 
the Weinstein conjecture and is similar to a result that he has established for the continuous
equation in an earlier paper\footnote{Weinstein's result makes 
significant use of the
pseudoconformal invariance of the continuous equation, which is 
unfortunately not present for the
discrete problem. In particular, for the decaying result, 
he imposes the condition $\norm{x
u_0}{L^2}<\infty$ on the initial data. In our results, 
it suffices to assume somewhat less - in
the one dimensional case: $\norm{|n|^{3/10}u_n}{l^{2,1}(\cz)}<<1$ and in the two dimensional case, one has to
impose $\norm{|n|^{1/2}u_n}{l^{2,1}(\cz^2)}<<1$, see the precise statement of Theorem 
\ref{theo:3} for conditions in terms of the $l^p$ spaces.}
\cite{wein1}, see also the excellent expository paper \cite{wein}.  
\begin{theorem}
\label{theo:3}
Let $\si> 2/d$ and $d\leq 2$. There exists an $\ve$, so that whenever \\
$\norm{u_n(0)}{l^{(8+2d)/(d+7)}}\leq \ve$, one has for  all 
$p: 2\leq p \leq (8+2d)/(d+1)$, 
\begin{equation}
\label{eq:152}
\norm{\{u_n(t)\}}{l^p}\leq C t ^{-d(p-2)/(3p)}\norm{u_n(0)}{p'}.
\end{equation}
which is the generic 
rate of decay for the free solutions (see \eqref{eq:151}  below). In particular, 
 no standing wave
solutions are possible under the smallness assumptions outlined above.
\end{theorem}
%PANOS IT WOULD BE GREAT IF YOU COULD TEST WHETHER YOU HAVE OR DON'T HAVE THE RATE OF DECAY CLAIMED
%HERE FOR $P>(8+2d)/(d+1)$. I HAVE THE FEELING THAT THIS SHOULD BE SHARP, i.e. you don't have this
%rate (especially in d=2), BUT I HAVE BEEN WRONG
%BEFORE MORE OFTEN THAT I WISH TO ADMIT.\\
%ALSO YOU SEE THE IMPORTANCE OF THE CONDITION $\si>2/d$ here pretty well.
\begin{proof} 
By interpolation between \eqref{eq:1} and \eqref{eq:2}, we obtain  
\begin{equation}
\label{eq:151}
\norm{\{e^{i t \De_d} u_n(0)\}}{l^{p}}\leq
C<t>^{-d(p-2)/(3p)}\norm{\{u_n(0)\}}{l^{p'}}.
\end{equation}
valid for all $2\leq p\leq \infty$. \\
Note that from
\eqref{eq:151}, we deduce that for every $p\geq 2$, $\norm{\{e^{i t \De_d}
u_n(0)\}}{l^{p}}\lesssim t^{-d(p-2)/(3p)}$. 

We would like to establish an a priori bound (which simultaneously implies existence) for the
solution $\{u_n\}$, that is we want to place it in 
$$
X=\{\{u_n\}: 
\norm{\{u_n(t)\}}{l^{p}}\leq 
2 C t^{-d(p-2)/(3p)}\norm{\{u_n(0)\}}{l^{p'}}\}.
$$
where $C$ is the constant in the 
Strichartz inequality.  
We have by the decay estimate \eqref{eq:151}
\begin{eqnarray*}
\norm{\{\La u_n(t)\}}{l^{p}}&\leq& C<t>^{-d(p-2)/(3p)} \norm{u_n(0)}{l^{p'}}+ \\
&+& C_1 \intl_0^t \f{1}{<t-s>^{d(p-2)/(3p)}}
\norm{u_n(s)}{l^{(2\si+1)p'}}^{2\si+1}ds.
\end{eqnarray*}
Since $(2\si+1)p'>(4/d+1)(8+2d)/(d+7)\geq (8+2d)/(d+1)=p$ for $d\leq 2$, 
 we have by the
inclusion $l^{p}\hookrightarrow l^{(2\si+1)p'}$
\begin{eqnarray*}
\norm{\{\La u_n(t)\}}{l^{p}} &\leq& C<t>^{-d(p-2)/(3p)} \norm{u_n(0)}{l^{p'}}+ \\
&+& C_1 \intl_0^t \f{1}{<t-s>^{d(p-2)/(3p)}}
\norm{u_n(s)}{l^{p}}^{2\si+1}ds\leq \\
&\leq & C<t>^{-d(p-2)/(3p)} \norm{u_n(0)}{l^{p'}}+\\
&+& C_2 \norm{u_n(0)}{l^{p'}}^{2\si+1} \intl_0^t \f{1}{<t-s>^{d(p-2)/(3p)}}
\f{1}{<s>^{d(p-2)(2\si+1)/(3p)}}ds.
\end{eqnarray*}
Note that $d(p-2)(2\si+1)/(3p)>d(p-2)(4/d+1)/(3p)>1$, and therefore the integral term above is
controlled by a constant times $t^{-d(p-2)/(3p)}$. As a consequence,
$$
\norm{\{\La u_n(t)\}}{l^{p}} \leq 2C <t>^{-d(p-2)/(3p)} \norm{u_n(0)}{l^{p'}},
$$
provided $C_3\norm{u_n(0)}{l^{p'}}^{2\si}\leq C_3\ve^{2\si}<C$.

The contractivity of the map $\La:X\to X$ follows in the same manner, given that $\ve<<1$.
A fixed point argument shows that a solution exists together with the estimate 
$\norm{\{u_n(t)\}}{l^p}\lesssim t^{-d(p-2)/(3p)}$.
\end{proof}

%%%%%%%%%%%%%%%%%%%%%%%%%%%%%%%%%%%%%%%

\begin{figure}[tbp]
\epsfxsize=8cm %\centerline{}
\centerline{\epsffile{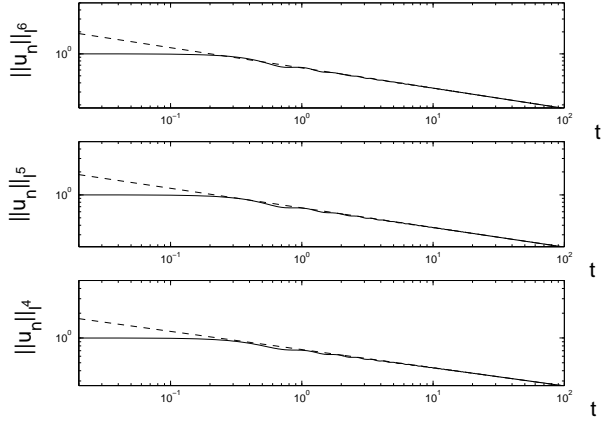}}
\caption{Log-log plot of the temporal evolution of the $l^{4}$ (bottom panel),
$l^5$ (middle panel) and $l^6$ (top panel) norms  
%norm 
in a 
nonlinear one dimensional 
Schr{\"o}dinger lattice. The intial condition contains one site excited
with $u_0=1$. The dashed line shows the best fits (for times 
$20 \leq t \leq 90$) that are given respectively by 
$||u_n||_{l^4} \sim t^{-0.221}$, $||u_n||_{l^5} \sim t^{-0.257}$
and $||u_n||_{l^6} \sim t^{-0.277}$.}
\label{dfig1a}
\end{figure}

%%%%%%%%%%%%%%%%%%%%%%%%%%%%%%%%%%%%%%%

The numerical simulations performed in a $d=1$, $\sigma=3$ lattice, with
an initial condition of a single excited site with $u_0=1$ showed 
(see Fig. \ref{dfig1a}) that
the actual decay rate is larger than that predicted by Theorem \ref{theo:3}.
In particular, for $p=4$ and $p=5$, the theorem predicts decay rates of
$t^{-1/6}$ and $t^{-1/5}$ respectively, while the numerical simulations
show decay rates of $t^{-0.221}$ and $t^{-0.257}$ respectively. For the 
case of $p=6$, the decay rate is faster and given by $t^{-0.277}$
(cf. with the theoretical prediction of $t^{-2/9}$).
%It is worth noting that when we consider multiple sites excited
%by the initial condition then, we get decay rates which are much
%closer to the ones theoretically predicted above. More specifically, if
%we excite 2 nodes (rather than a single one), with 3 sites betwen them,
%then the decay rates are $t^{-0.176}$, $t^{-195}$ and $t^{-0.203}$.

\section{Applications to the discrete Klein-Gordon equation}
For the discrete Klein-Gordon equation, 
as mentioned above we consider nonlinearities of the 
form $|u|^{2\si} u$. We can deduce essentially the same results as in the case for the discrete nonlinear Schr\"odinger equation. We state them without proofs, since they follow by exactly the same proofs as in the Schr\"odinger case (by the exact same decay and Strichartrz estimates).
\begin{theorem}
\label{theo:20}
Let $\si\geq 3$. 
There exists an $\ve>0$ and a constant $C$, so that whenever  
$\norm{\{u_n(0)\}}{l^2}, \norm{\{\p_t u_n(0)\}}{l^2}\leq \ve$ 
there exists a unique global  solution to the one-dimensional discrete 
Klein-Gordon equation with a nonlinearity $|u|^{2\si} u$, satisfying
$$
\norm{\{u_n(t)\}}{L^ql^r}\leq C\ve,
$$
for all Strichartz admissible pairs $(q,r)$. In particular, 
$\norm{u_n(t)}{l^r}\to 0$ as $t\to\infty$ for every $r>2$.
\end{theorem}
In the next theorem, we establish the Weinstein excitation 
threshold conjecture in the Klein-Gordon case. 
\begin{theorem}
\label{theo:30}
Let $\si> 2$. There exists an $\ve$, so that whenever 
$\norm{u_n(0)}{l^{5/4}}\leq \ve$, $ \norm{\p_t 
u_n(0)}{l^{5/4}}\leq \ve$, one has for  all 
$p: 2\leq p \leq 5$, 
\begin{equation}
\label{eq:1520}
\norm{\{u_n(t)\}}{l^p}\leq C t ^{-(p-2)/(3p)}\norm{u_n(0)}{p'}.
\end{equation}
In particular, there are no small standing wave solutions to the discrete Klein-Gordon equation.
\end{theorem}

%%%%%%%%%%%%%%%%%%%%%%%%%%%%%%%%%%%%%%%

\begin{figure}[tbp]
\epsfxsize=8cm %\centerline{}
\centerline{\epsffile{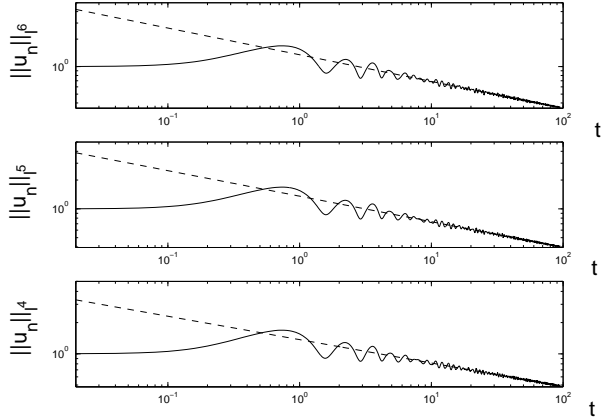}}
\caption{Same as the previous figure but for the nonlinear Klein-Gordon
lattice.
%Log-log plot of temporal evolution of the $l^{4}$ (bottom panel),
%$l^5$ (middle panel) and $l^6$ (top panel) norms  
%%norm 
%in a 
%linear one dimensional lattice. The intial condition contains one site excited
%with $u_0=1$. 
The dashed line shows the best fits (for times 
$20 \leq t \leq 90$) that are given respectively by 
$||u_n||_{l^4} \sim t^{-0.226}$, $||u_n||_{l^5} \sim t^{-0.267}$
and $||u_n||_{l^6} \sim t^{-0.292}$.}
\label{dfig1c}
\end{figure}

%%%%%%%%%%%%%%%%%%%%%%%%%%%%%%%%%%%%%%%

The numerical results, in this case as well, show a faster decay than
theoretically predicted. Furthermore, the decay is slightly faster in
this model for the same norms in comparison with the corresponding
cases for the Schr{\"o}dinger equation. More specifically, in 
the Klein-Gordon model, $||u_n||_{l^4} \sim t^{-0.226}$, 
$||u_n||_{l^5} \sim t^{-0.267}$ and $||u_n||_{l^6} \sim t^{-0.292}$.
It is worth noting, however, that when we consider multiple sites excited
by the initial condition, then we obtain decay rates which are much
closer to the ones theoretically predicted above. More specifically, if
we excite 2 nodes (rather than a single one), with 3 sites betwen them,
then the decay rates are $t^{-0.207}$, $t^{-245}$ and $t^{-0.270}$
which are considerably 
closer to the theoretically predicted exponents of $1/6$, 
$1/5$ and $2/9$ respectively.

\section{Appendix}
\begin{proof}(Lemma \ref{le:1}) we present the proof in the $d$ dimensional case. \\
We use the representation of $h(\sqrt{1-\De_d})$ to conclude that it suffices to show that 
for every $N$, there exists $C_N$, so that 
\begin{equation}
\label{eq:90}
|b_m|:=|\intl_{[0,1]^d} h(1+4\suml_{j=1}^d \sin^2(\pi
k_j)) e^{2\pi i m\cdot k}dk|\leq C_N <m>^{-N}.
\end{equation}
Indeed, if we had \eqref{eq:90}, then by the expression  $(h(\sqrt{1-\De_d})u)_n=\sum_{m} b_{n-m} u_m$, we conclude 
$$
\norm{h(\sqrt{1-\De_d})u}{l^p}\lesssim  \norm{b}{l^1} \norm{u}{l^p}\leq C_{d} \norm{<\cdot>^{-d+2}}{l^1}  \norm{u}{l^p}.
$$
For the proof of \eqref{eq:90}, use integration by parts and  the fact that boundary terms disappear (due to the fact that $h(1+4\suml_{j=1}^d \sin^2(\pi
k_j)) $ is one-periodic in every variable $k_j$). We get  
$$
b_m=\f{c}{m_j} \int_{[0,1]^d} h'(1+4\suml_{j=1}^d \sin^2(\pi
k_j))\sin(2\pi k_j)e^{2\pi m\cdot k} dk.
$$
Clearly, one keeps integrating by parts to obtain arbitrary large rate of decay in all variables $m_j$. The lemma is proven.
\end{proof}

\end{document}